\documentclass[aps,pra,floatfix,reprint]{revtex4-1}

\usepackage{graphicx}
\usepackage{bm}
\usepackage{amssymb}
\usepackage{rotating}
\usepackage{amsmath}
\usepackage[caption=false]{subfig}

\newcommand{\sub}[1]{\ensuremath{_{\textrm{#1}}}}

\begin{document}

\title{Generalized Pauli Conditions on the Spectra of One-electron Reduced Density Matrices of Atoms and Molecules}

\author{Romit Chakraborty and David A. Mazziotti}
\email{damazz@uchicago.edu}

\affiliation{Department of Chemistry and The James Franck Institute, The University of Chicago, Chicago, IL 60637}%

\date{Submitted January 23, 2014; revised March 31, 2014}

\begin{abstract}

  The Pauli exclusion principle requires the spectrum of the occupation numbers of the one-electron reduced density matrix (1-RDM) to be bounded by one and zero.  However, for a 1-RDM from a wave function there exist additional conditions on the spectrum of occupation numbers known as pure $N$-representability conditions or generalized Pauli conditions. For atoms and molecules we measure through a Euclidean-distance metric the proximity of the 1-RDM spectrum to the facets of the convex set (polytope) generated by the generalized Pauli conditions. For the ground state of any spin symmetry, as long as time-reversal symmetry is considered in the definition of the polytope, we find that the 1-RDM's spectrum is pinned to the boundary of the polytope. In contrast, for excited states we find that the 1-RDM spectrum is not pinned. Proximity of the 1-RDM to the boundary of the polytope provides a measurement and classification of electron correlation and entanglement within the quantum system.  For comparison, this distance to the boundary of the generalized Pauli conditions is also compared to the distance to the polytope of the traditional Pauli conditions, and the distance to the nearest 1-RDM spectrum from a Slater determinant.  We explain the difference in pinning in the ground- and excited-state 1-RDMs through a connection to the $N$-representability conditions of the two-electron reduced density matrix.

\end{abstract}

\maketitle

\section{Introduction}

The {\em Pauli exclusion principle} states that two identical fermions cannot occupy the same quantum state~\cite{P25}. Postulated by Pauli in 1925 to explain atomic transitions~\cite{S24}, this principle plays a key role in predicting electronic configurations of atoms and molecules.  Stated otherwise, the Pauli principle says that the fermion occupation numbers $\lambda_i$ of a quantum system must lie between $0$ and $1$
\begin{equation}
0 \le \lambda_i \le 1 .
\end{equation}
Subsequent work by Dirac~\cite{D26} and Heisenberg~\cite{H26} showed that this principle arises from the antisymmetry of the fermion wave function.

As discussed by von Neumann~\cite{VN}, a general $N$-fermion quantum state is expressible by an $N$-fermion ensemble density matrix
\begin{equation}
^{N} D(1,2,..,N;{\bar 1},{\bar 2},..,{\bar N}) = \sum_{i}{ w_{i} \Psi_{i}(1,2,..,N) \Psi_{i}^{*}({\bar 1},{\bar 2},..,{\bar N}) }
\end{equation}
where $w_{i}$ are non-negative weights that sum to unity, $\Psi_{i}(1,2,..,N)$ are $N$-fermion wave functions, and each number denotes the spatial and spin coordinates of a fermion. Integration of the $N$-fermion {\em ensemble density matrix} over the coordinates of all fermions save one yields the one-fermion reduced density matrix (1-RDM)
\begin{equation}
^{1} D(1;{\bar 1}) = \int{ ^{N} D(1,2,..,N;{\bar 1},2,..,N) d2d3..dN } .
\end{equation}
Like the $N$-fermion density matrix, the 1-RDM must be (i) Hermitian, (ii) normalized, and (iii) positive semidefinite. However, the 1-RDM must also obey additional constraints to ensure that it is derivable from the integration of an $N$-fermion ensemble density matrix $^{N} D$.  These additional constraints are known as {\em ensemble $N$-representability conditions}~\cite{C63}.  The eigenfunctions of the 1-RDM are known as {\em natural orbitals} while the eigenvalues of the 1-RDM are known as the {\em natural occupation numbers}. Coleman showed that the Pauli exclusion principle applied to the natural occupation numbers imposes necessary and sufficient ensemble $N$-representability conditions on the 1-RDM, that the eigenvalues of the 1-RDM must lie between 0 and 1~\cite{C63}.

\begin{figure}[t!]

  \includegraphics[scale=0.42]{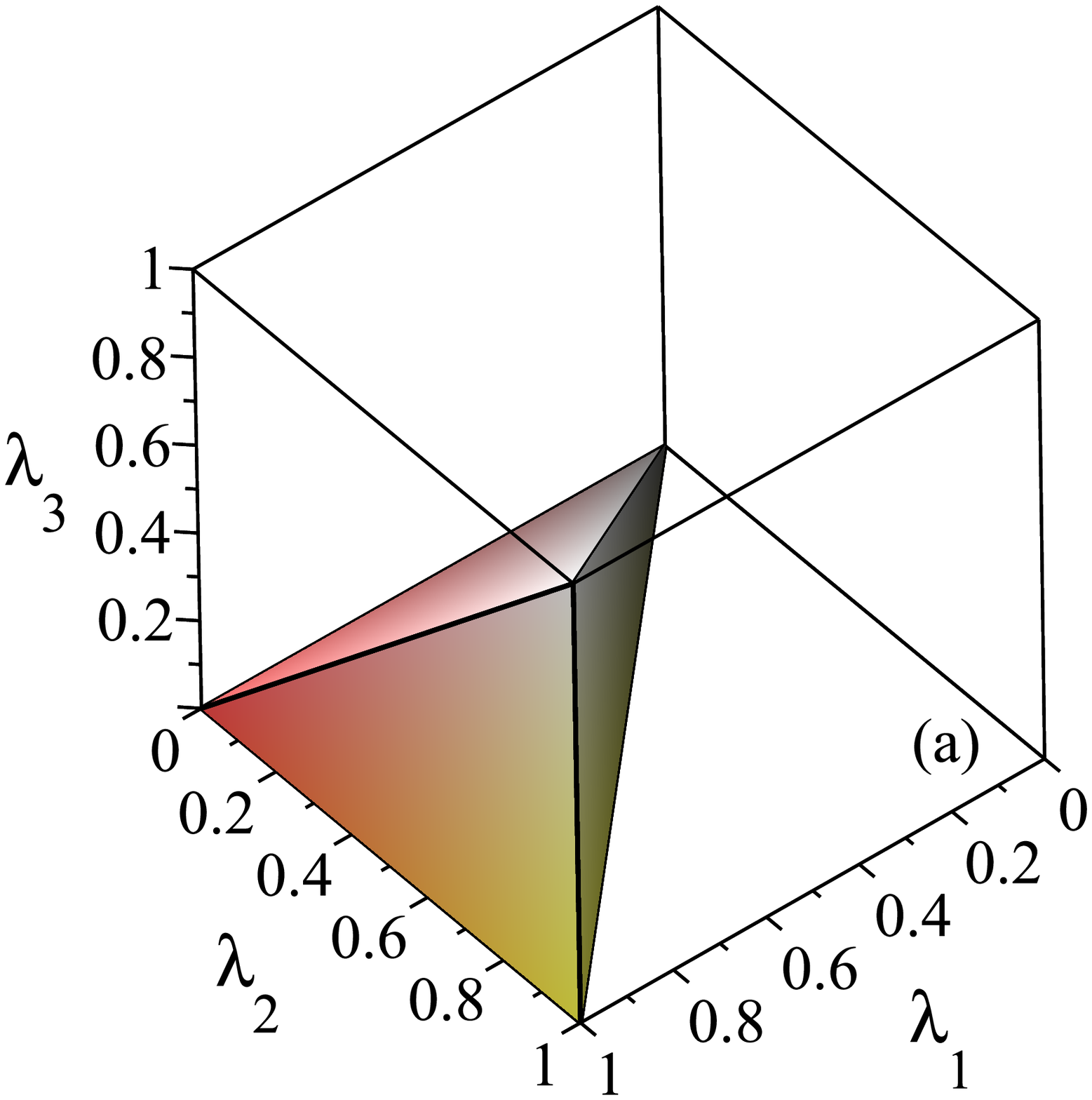}

  \vspace{0.5cm}

  \includegraphics[scale=0.42]{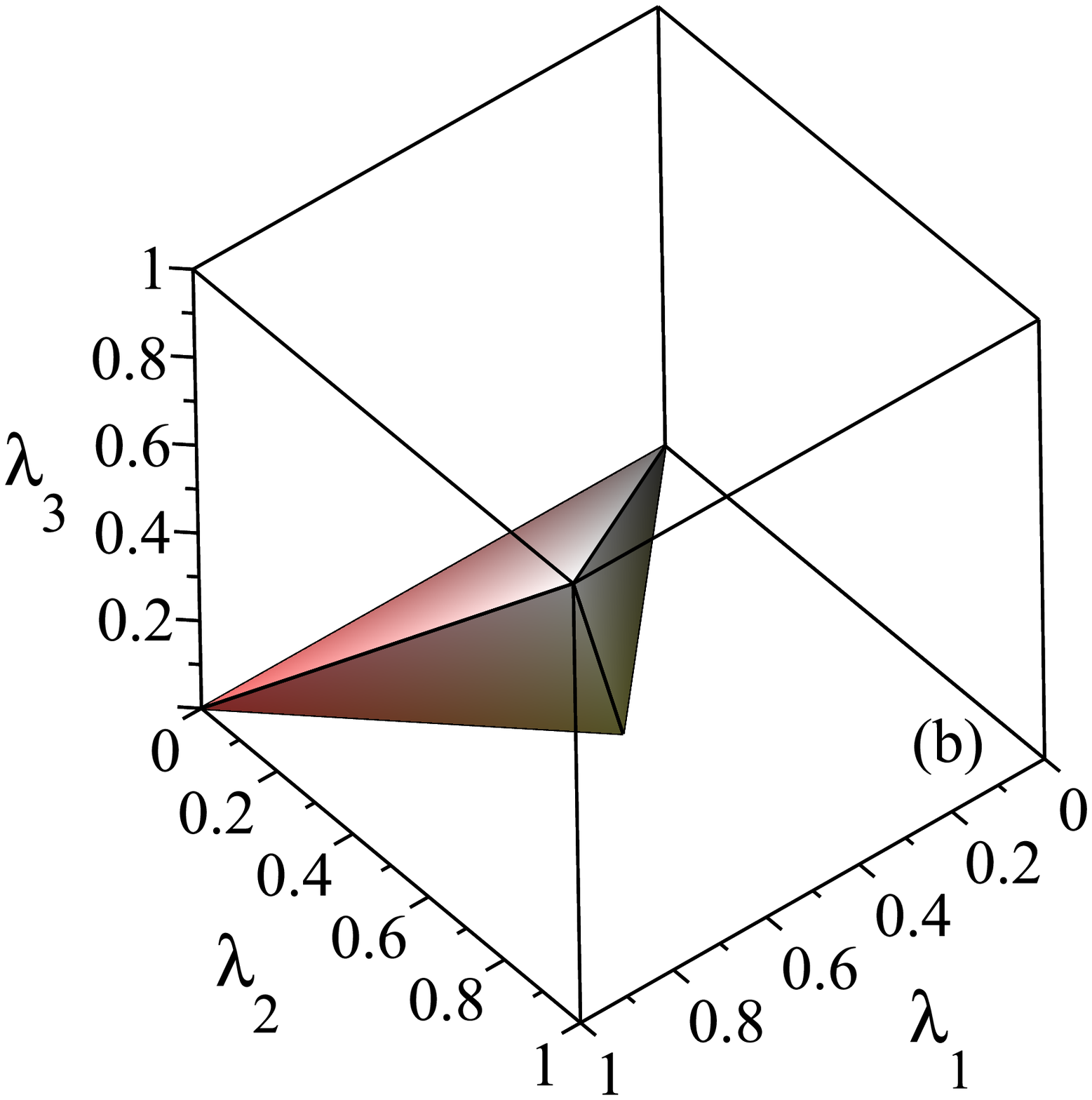}

  \caption{(Color online) The sets of (a) ensemble and (b) pure $N$-representable 1-RDMs are shown for a general three-electron ($N=3$) and six-orbital ($r=6$) quantum system.  The plane defined by the Borland-Dennis equalities in Eq.~(\ref{eq:BD2}) causes the pure $N$-representable set of 1-RDMs in (b) to be significantly smaller than the ensemble $N$-representable set of 1-RDMs in (a).  The sets are shown in terms of the first three natural occupation numbers, $\lambda_{1}$, $\lambda_{2}$, and $\lambda_{3}$, ordered from highest to lowest, relative to a fixed set of natural orbitals.  These three occupation numbers provide a complete three-dimensional description of the pure 1-RDM spectra because the other occupation numbers are determined from the Borland-Dennis equalities in Eq.~(\ref{eq:BD1}); they provide a partial description of the full five-dimensional ensemble 1-RDM spectra.}   \label{fig:genp}
\end{figure}

While the Pauli conditions of the 1-RDM are complete ensemble $N$-representability conditions, additional conditions on the 1-RDM are required to ensure that it arises from the integration of an $N$-fermion {\em pure density matrix}
\begin{equation}
^{N} D(1,2,..,N;{\bar 1},{\bar 2},..,{\bar N}) = \Psi(1,2,..,N) \Psi^{*}({\bar 1},{\bar 2},..,{\bar N})
\end{equation}
where the $^{N} D$ can be spectrally resolved in terms of the single $N$-fermion wave function $\Psi({\bar 1},{\bar 2},..,{\bar N}) $.  These additional 1-RDM constraints are known as {\em pure $N$-representability conditions} or {\em generalized Pauli conditions}~\cite{C63, S66, BD72, K06, PRL13, PRA13}.  The pure $N$-representability conditions of the 1-RDM depend only on its natural occupation numbers~\cite{C63}, and hence, we will use the terms $N$-representability of the 1-RDM and $N$-representability of the 1-RDM spectrum, interchangeably.  Smith showed that pairwise degeneracy of occupation numbers are sufficient to ensure pure $N$-representability of the 1-RDM~\cite{S66}.  Furthermore, he showed that such degeneracy occurs naturally in even-$N$ quantum systems with time-reversal symmetry.  In 1972 Borland and Dennis reported pure $N$-representability conditions for active space of three fermions in six orbitals denoted by $\wedge^3[\mathcal{H}_6]$, on the basis of numerical calculations~\cite{BD72}. For an ordered set of occupation numbers $\lambda_i \geq \lambda_{i+1}$, their conditions are given by
\begin{align}
\label{eq:BD1}
\lambda_1 + \lambda_6 = \lambda_2 + \lambda_5 = \lambda_3 + \lambda_4 = 1 \\
\label{eq:BD2}
\lambda_5 + \lambda_6 - \lambda_4 \geq 0 .
\end{align}
Until recently, a systematic enumeration of generalized Pauli constraints has been elusive.  Based on work in quantum marginal theory, Klyachko was able to list the necessary and sufficient constraints for $N$ fermions in $r$ orbitals~\cite{K06,AK08}. These constraints are expressible in the form of linear inequalities in the occupation numbers $\{\lambda_i\}$
\begin{equation}
\kappa_{0} + \kappa_{1}\lambda_{1} + \cdots + \kappa_{r}\lambda_{r}\geq 0,\label{eq:Klchko}
\end{equation}
which can be visualized as a convex polytope in $\mathbb{R}^r$~\cite{R70}. The polytope for $N=3$ and $r=6$, whose boundary is defined by the Borland-Dennis inequalities in Eqs.~(\ref{eq:BD1}) and~(\ref{eq:BD2}), is shown in Fig.~\ref{fig:genp}.

In this paper we study the generalized Pauli conditions for ground and excited states of three- and four-electron atoms and molecules including Li, LiH, BH, and BeH$_{2}$ as well as H$_{3}$ and H$_{4}$.   Both H$_{3}$ and H$_{4}$ are studied at both equilibrium and non-equilibrium geometries.  Previous work has explored these conditions for the ground states of a quantum harmonic oscillator in a harmonic potential~\cite{PRL13} and the lithium isoelectronic sequence~\cite{PRA13}.  While previous work examined the residual of the generalized Pauli conditions to measure saturation of the inequalities~\cite{PRL13,PRA13}, we measure the distance of the 1-RDM from the boundary of the set of pure $N$-representable 1-RDMs through a Euclidean metric~\cite{H78}.  Specifically, we compute the minimum Euclidean distance between the natural occupation numbers of the 1-RDM to the boundary of the polytope generated by the generalized Pauli conditions.  The distance of the 1-RDM spectrum to the nearest facet of the polytope provides a measurement and classification of electron correlation and entanglement.  For comparison we also compute the minimum Euclidean distance to the boundary of ensemble $N$-representable 1-RDMs and the minimum Euclidean distance to the nearest non-interacting 1-RDM, that is a 1-RDM corresponding to a Slater determinant.  The computations of both the 1-RDMs and the Euclidean distances are performed in the computer algebra program Maple~\cite{Maple} with arbitrary-precision floating-point arithmetic.  The 1-RDMs are computed from full configuration interaction (FCI) calculations; in several cases for comparison, the 1-RDMs are also computed from variational 2-RDM computations~\cite{EJ00, NNE01, M02, ZBF04, M04, *M06, CSL06, GM06, HM06, RM08, NBF08, VVV09, S10, GM10, PGG11, M11, B12} with approximate ensemble $N$-representability conditions on the 2-RDM~\cite{GP64, E78, CY00, RDM, M12}.

For three-electron atoms and molecules we find that the 1-RDM of the ground-state wave function is always pinned to the boundary of the pure $N$-representable set of 1-RDMs.   Importantly, this pinning of the 1-RDM to the boundary of its pure $N$-representable set occurs even for strongly correlated three-electron molecules like equilateral H$_{3}$.  This constitutes the first numerical evidence of pinning in a strongly correlated molecule.  For four-electron atoms and molecules we find that the 1-RDM of the ground-state wave function is not pinned to the boundary of the generalized Pauli conditions derived by Klyachko~\cite{K06}.  However, because these four electron atoms molecules obey time-reversal symmetry, we should in fact consider the generalized Pauli conditions derived by Smith~\cite{S66} that explicitly account for this additional symmetry.   With time-reversal symmetry included the ground-state 1-RDM of four-electron systems is found in all cases to be pinned to the boundary of the generalized Pauli conditions.  In contrast to the ground states, some of the 1-RDMs from excited states of a given spin symmetry are found to be significantly inside the set of pure $N$-representable 1-RDMs.  In section~\ref{sec:D2} for both ground and excited states we theoretically motivate these computational findings through an analysis of the pure $N$-representable set of 2-RDMs, the set of 2-RDMs that correspond to at least one $N$-fermion wave function.

\section{Pinning of 1-RDM Spectra}

In section~\ref{sec:D2} we discuss a necessary 2-RDM condition for pinning of the 1-RDM to the boundary of pure $N$-representable 1-RDMs.  The condition is valid for a one-electron basis set in the case of a finite rank $r$ as well as in the limit that the rank $r$ approaches infinity.  In section~\ref{sec:ED} we compute the minimum Euclidean distances to the boundaries of the sets of $N$-representable 1-RDMs that are pure and ensemble, respectively.  These Euclidean distances are useful for both measuring and classifying electron correlation and entanglement.  Finally, in section~\ref{sec:TR} we discuss the pure $N$-representable 1-RDM set for even-$N$ quantum systems with time-reversal symmetry.

\subsection{Necessary 2-RDM condition}

\label{sec:D2}

The ground-state energy of an $N$-electron system can be expressed as a functional of the 2-RDM
\begin{equation}
E = \frac{N(N-1)}{2} {\rm Tr}(^{2} K \, ^{2} D)
\end{equation}
where $^{2} D$ is the 2-RDM
\begin{equation}
^{2} D(1,2;{\bar 1},{\bar 2}) = \int{ ^{N} D(1,2,..,N;{\bar 1},{\bar 2},..,N) d3..dN }
\end{equation}
and $^{2} K$ is the two-electron reduced Hamiltonian
\begin{equation}
^{2} K(1,2;{\bar 1},{\bar 2}) = \frac{1}{N-1} \sum_{i=1}^{2}{(-\frac{1}{2} {\hat \nabla}^{2}_{i} - \sum_{k}{\frac{Z_{k}}{r_{ik}}} )} + \frac{1}{r_{12}}.
\end{equation}
Minimization of the energy over the convex set $E_{N}^{2}$ of ensemble $N$-representable 2-RDMs yields the ground-state energy $E_{0}$ of the $N$-electron quantum system~\cite{C63, CY00, RDM}
\begin{equation}
E_{0} = \min_{^{2} D \in E_{N}^{2}}{ E(^{2} D) } .
\end{equation}
Because the energy is a linear functional of the 2-RDM, the optimal 2-RDM for a non-degenerate ground state lies on the boundary of the convex set $E_{N}^{2}$.  

A 2-RDM that is {\em ensemble $N$-representable} must be derivable from the integration of at least one $N$-electron density matrix.  The 2-RDM that is {\em pure $N$-representable} must also be derivable from the integration of at least one pure $N$-electron density matrix.  From these definitions it follows that the set $P_{N}^{2}$ of pure $N$-representable 2-RDMs is contained in the set $E_{N}^{2}$ of ensemble $N$-representable 2-RDMs, that is $P_{N}^{2} \subset E_{N}^{2}$.  By the energy minimization discussed above, the 2-RDM of a non-degenerate ground state lies on the boundary of the ensemble set $E_{N}^{2}$.  Because the 2-RDM of a non-degenerate ground state is pure $N$-representable and $P_{N}^{2} \subset E_{N}^{2}$, it also lies on the boundary of the pure set $P_{N}^{2}$.  In contrast, an excited-state 2-RDM generally lies {\em inside} the ensemble set $E_{N}^{2}$ of 2-RDMs.

Because a 1-RDM arises from the integration of a ground-state 2-RDM over the coordinates for electron two, the ground-state 1-RDM can lie in the boundary of its pure $N$-representable set $P_{N}^{1}$ only if it derives from a 2-RDM that lies on the boundary of its pure $N$-representable set $P_{N}^{2}$. Hence, the 2-RDM contains a necessary condition for the pinning of the 1-RDM spectra to the generalized Pauli conditions.  This result also provides important information about the potential difference in pinning of the ground-state and excited-state 1-RDM spectra.  Because the ground-state 2-RDM lies on the boundary of the ensemble $N$-representable 2-RDM set $E_{N}^{2}$ and hence, on the boundary of the pure $N$-representable 2-RDM set $P_{N}^{2}$, it is possible for the ground-state 1-RDM to lie on the boundary of the pure $N$-representable 1-RDM set $P_{N}^{1}$.  In contrast, because an excited-state 2-RDM does not necessarily lie on the boundary of $E_{N}^{2}$ or $P_{N}^{2}$, it is possible for the excited-state 1-RDM (of a given spin symmetry) to lie in the interior (not on the boundary) of the set $P_{N}^{1}$.

\subsection{Euclidean distances}

\label{sec:ED}

In this section we develop optimization programs for computing the minimum Euclidean distance from a given 1-RDM's  the $r$-dimensional spectrum of natural occupation numbers $\vec{n} = \{\lambda_i\}$ to three other points in the Euclidean space of spectra: (i) the nearest point $\vec{p}$ on the boundary of the set of {\em pure} $N$-representable 1-RDMs, (ii) nearest point $\vec{e}$ on the boundary of the set of {\em ensemble} $N$-representable 1-RDMs, and (iii) the nearest point $\vec{s}$ corresponding to a 1-RDM with a Slater determinant pre-image.   These three distances are useful in assessing a quantum system's electron correlation as well as its purity.  A quantum system is pure if and only if it is described by a single wave function rather than an ensemble of wave functions.

Firstly, we can compute the minimum distance of the spectrum $\vec{n}$ to the boundary of {\em pure} $N$-representability 1-RDMs or in other words the minimum distance of $\vec{n} $ to the polytope facets defined by the generalized Pauli conditions $M\vec{p} \ge b$, the Pauli conditions $0 \le p_i \le 1$, the trace condition $\sum_{i}{ p_{i} } = N$, and a condition ordering the occupation numbers from highest to lowest in magnitude $p_{i+1} \le p_{i}$ where $\vec{p}$ is any point in the $d$-dimensional Euclidean space $\mathbb{R}^r$:
\begin{eqnarray}
\min_{j}{ \min_{\vec{p} \in \mathbb{R}^r}{||\vec{n} - \vec{p}||} } \label{eq:p1} \\
{\rm such~that} \hspace{0.2cm} \sum_{i}{ p_{i} } & = & N \\
p_{i+1} & \le & p_{i}~{\rm for~all}~i \in [1,r-1] \\
0 \le & p_i & \le 1~~{\rm for~all}~i \in [1,r]\\
M\vec{p} & \ge & {b} \\
\sum_{i}{M^{j}_{i} p_{i}} & = & b_{j} \label{eq:p2}
\end{eqnarray}
The boundary of the convex polytope defined by the affine inequalities $M\vec{p} \ge b$ is the union of the hyperplanes defined by the saturated inequalities intersected with the domain under consideration.  A point lies on the boundary when at least one of the constraints $M\vec{p} \ge {b}$ is saturated.  If we name the saturated constraint $j$, then the constraints in Eqs.~(\ref{eq:p1}-\ref{eq:p2}) express that $\vec{p}$ belongs to the facet $j$ of the convex polytope.  The algorithm works by (i) minimizing the Euclidean distance to each facet $j$ and (ii) minimizing over the results from (i).  For $N=3$ the constraints $M\vec{p} \ge {b}$ represent the Borland-Dennis constraints, and for $N=4$ the constraints represent either the Klyachko or Smith constraints.  We shall refer to the set of pure $N$-representable 1-RDMs as the {\em pure set} and the minimum Euclidean distance to the boundary of the pure set as the {\em pure distance}.

Secondly, we can compute the minimum Euclidean distance to the boundary of the set of ensemble $N$-representable 1-RDMs, defined by the Pauli principle and the trace condition, as follows:
\begin{eqnarray}
\min_{j}{ \min_{b \in \{0,1\}}{ { \min_{\vec{e} \in \mathbb{R}^r}{||\vec{n} - \vec{e}||} }}} \\
{\rm such~that} \hspace{0.2cm} \sum_{i}{ e_{i} } & = & N \\
e_{i+1} & \le & e_{i}~{\rm for~all}~i \in [1,r-1] \\
0 \le & e_i & \le 1~~{\rm for~all}~i \in [1,r] \\
e_{j} & = & b
\end{eqnarray}
where $\vec{e} = \{e_i\}$ is any point in the {\em ensemble set}.   The saturation of the constraints corresponds to an element with one occupation number being either 0 or 1.  We shall refer to the minimum Euclidean distance to the boundary of the ensemble set as the {\em ensemble distance}.  The pure distance is less than or equal to the ensemble distance because $P^{1}_{N} \subset E^{1}_{N}$.

Finally, the natural occupation numbers of a non-interacting 1-RDM, the 1-RDM that derives from a Slater determinant, are either fully occupied or empty.  These eigenvalues defined the components of a Slater point $\vec{s} $ in Euclidean space
\begin{equation}
\vec{s} = (1, 1, \cdots , 0,0) .
\end{equation}
The minimum Euclidean distance from a given 1-RDM's spectrum $\vec{n}$ to the nearest Slater point can be computed from the following minimization:
\begin{equation}
\min_{\vec{s} \in \mathbb{S}^r}{||\vec{n} - \vec{s}||}
\end{equation}
where $\mathbb{S}^r$ denotes the set of all Slater points.  The non-convex set of Slater points is a subset of the points on the boundary of the ensemble $N$-representable set $E^{1}_{N}$.  Hence, the minimum distance to the boundary of the ensemble set (ensemble distance) is strictly less than or equal to the minimum distance to the nearest Slater point.  The Euclidean distance $||\vec{n}-\vec{s}||$ of the spectrum from the nearest Slater point, which we call the {\em Slater distance}, gives a useful measure of electron entanglement and correlation that equals zero in the absence of correlation.

\subsection{Time-reversal symmetry}

\label{sec:TR}

Smith proved two key results in the study of the pure $N$-representability conditions of the 1-RDM~\cite{S66}.  Firstly, for an even-$N$ quantum state, if all of the eigenvalues of the state's 1-RDM are evenly degenerate, then the 1-RDM is pure $N$-representable.  Secondly, if an even-$N$ quantum state has time-reversal symmetry, then all of the eigenvalues of the state's 1-RDM are evenly degenerate.  Hence, {\em if an even-$N$ quantum state has time-reversal symmetry, the degeneracy of the eigenvalues of the 1-RDM is necessary and sufficient for the 1-RDM to be pure $N$-representable.}

Smith's set of pure $N$-representable 1-RDMs with time-reversal symmetry $S_{N}^{1}$ is a subset of the pure $N$-representable set of 1-RDMs $P_{N}^{1}$, that is $S_{N}^{1} \subset P_{N}^{1}$.  Importantly, any 1-RDM from a pure state with time-reversal symmetry is pinned to the boundary of the Smith set $S_{N}^{1}$.  For example, for $N = 4$ and $r = 8$ the Smith set is characterized by the four equalities between natural occupation numbers
\begin{eqnarray}
\lambda_{1} & = & \lambda_{2} \\
\lambda_{3} & = & \lambda_{4} \\
\lambda_{5} & = & \lambda_{6} \\
\lambda_{7} & = & \lambda_{8} .
\end{eqnarray}
Each equality can be viewed as two inequalities; for example, the first equality can be expressed as the following two inequalities
\begin{eqnarray}
\lambda_{1} & \le & \lambda_{2} \\
\lambda_{2} & \le & \lambda_{1} .
\end{eqnarray}
Because these inequalities are always saturated, any 1-RDM in the Smith set is pinned to the boundary of the Smith set.  More generally, this result is true for any even $N$ and $r$.

For $N = 4$ and $r = 8$ the generalized Pauli inequalities on the natural occupation numbers of the 1-RDM, determined computationally by Borland and Dennis~\cite{BD72} and derived by Klyachko~\cite{K06}, are
\begin{eqnarray}
\label{eq:4a}
4-\sum_{i=1}^{8}{ M^{j}_{i} \lambda_{i} } & \ge & 0 \\
\label{eq:4b}
4+\sum_{i=1}^{8}{ M^{j}_{9-i} \lambda_{i} } & \ge & 0 .
\end{eqnarray}
Seven sets of coefficients $M^{j}_{i}$ are given in Table~\ref{t:coef} for a total of 14 inequalities.  When time-reversal symmetry is imposed on the 1-RDM by forcing its eigenvalues to be evenly degenerate, these generalized Pauli inequalities reduce to the traditional Pauli exclusion principle. Consequently, a 1-RDM with time-reversal symmetry is pinned to the Klyachko inequalities if and only if it is pinned to the traditional Pauli conditions, meaning the boundary of the ensemble $N$-representable 1-RDM set.  The 1-RDM with time-reversal symmetry has a spectrum that is typically not pinned to the convex Klyachko set $P_{N}^{1}$ because $P_{N}^{1}$ contains additional 1-RDMs than the convex Smith set $S_{N}^{1}$ that break time-reversal symmetry.

\begin{table}[t!]
  \caption{For $N = 4$ and $r = 8$ the generalized Pauli inequalities on the natural occupation numbers of the 1-RDM, shown in Eqs.~\ref{eq:4a} and~\ref{eq:4b}, have seven sets of coefficients $M^{j}_{i}$ displayed below where $j$ labels the set and $i$ labels the coefficient within a set.}
  \label{t:coef}

  \begin{ruledtabular}
  \begin{tabular}{ccccccccc}
    & \multicolumn{8}{c}{Coefficients in Eqs.~(\ref{eq:4a}) and~(\ref{eq:4b})} \\ \cline{2-9}
  j & $M^{j}_{1}$ & $M^{j}_{2}$ & $M^{j}_{3}$ & $M^{j}_{4}$ & $M^{j}_{5}$ & $M^{j}_{6}$ & $M^{j}_{7}$ & $M^{j}_{8}$ \\ \hline
  1 & 5 & -3 &  1 &  1 &  1 &  1 & -3 & -3 \\
  2 & 5 &  1 & -3 &  1 &  1 & -3 &  1 & -3 \\
  3 & 5 &  1 &  1 & -3 &  1 & -3 & -3 &  1 \\
  4 & 5 &  1 &  1 & -3 & -3 &  1 &  1 & -3 \\
  5 & 1 &  5 &  1 & -3 &  1 & -3 &  1 & -3 \\
  6 & 1 &  1 &  5 & -3 &  1 &  1 & -3 & -3 \\
  7 & 1 &  1 &  1 &  1 &  5 & -3 & -3 & -3
  \end{tabular}
  \end{ruledtabular}
\end{table}

\section{Applications}
\label{sec:app}

We evaluate the deviation of the 1-RDM spectrum in atoms and molecules from the boundary of the ensemble $N$-representable 1-RDM set and boundaries of the pure $N$-representable 1-RDM sets both with and without time-reversal symmetry.

\subsection{Computational Details}

For each atom or molecule the 1-RDM's spectrum of occupation numbers was obtained by computing the wave function from a full configuration interaction (FCI) calculation in arbitrary-precision floating-point arithmetic.  For comparison the 1-RDMs of several four-electron molecules were also computed without the wave function from the variational 2-RDM method~\cite{EJ00, NNE01, M02, ZBF04, M04, *M06, CSL06, GM06, HM06, RM08, NBF08, VVV09, S10, GM10, PGG11, M11, B12} with approximate ensemble $N$-representability conditions on the 2-RDM~\cite{GP64, E78, CY00, RDM, M12}.  The pure, ensemble and Slater distances were calculated in arbitrary-precision arithmetic by the constrained optimizations described in section~\ref{sec:ED}.  The FCI calculations computed all of the Hamiltonian’s eigenvalues through a QR method~\cite{QR}.  The distance calculations were performed with the sequential quadratic programming algorithm available in Maple~\cite{Maple}.  Calculations with 50 decimals of floating-point precision employed an optimality tolerance of $10^{-36}$.  The initial guess was selected randomly. All molecules were treated in the Slater-type-orbital (STO-3G) basis set in which each Slater function is expanded in three Gaussian functions~\cite{HSP69}.  The number $r$ of orbitals is always set to be twice the number $N$ of electrons.  To achieve either $N=3$ and $N=4$ with $2N$ orbitals, we froze core and virtual orbitals in atoms and molecules, as needed.  Electron integrals from the GAMESS package~\cite{G93} were employed. Molecular equilibrium geometries were obtained from the {\em Computational Chemistry Comparison and Benchmark Database}~\cite{CCCBD13}.

\subsection{Lithium}

\begin{figure}[t!]

  \includegraphics[scale = 0.7]{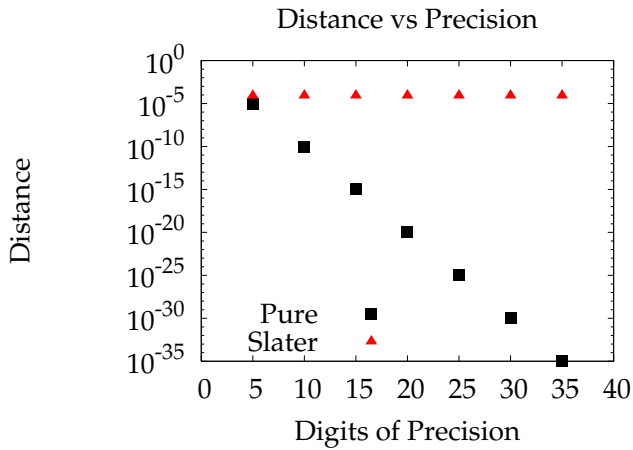}

  \caption{(Color online) For the ground state of the lithium atom the logarithm of the pure distance decreases linearly with the precision of the floating-point calculations.  The plot demonstrates that the ground-state 1-RDM spectrum of lithium is pinned to a facet to at least 35~digits of floating-point precision.}

  \label{fig:lip}

\end{figure}

For the three-electron lithium atom the Euclidean distances of the 1-RDM's spectrum to the pure and ensemble boundaries and the nearest Slater point were computed. Previous calculations on the lithium atom were inconclusive about whether the 1-RDM's spectrum was pinned or only nearly pinned (quasi-pinned) to one of the generalized Pauli constraints~\cite{PRA13}.  To resolve the issue, we performed both the FCI and the Euclidean-distance calculations with high-precision floating-point arithmetic {\em with as many as 35 digits of precision}.  Table~\ref{t:lip} shows the pure and Slater distances as functions of the floating-point precision.  While the Slater distance remains constant at 8.53 x $10^{-5}$ as the precision is increased, the logarithm of the pure distance decreases linearly with the precision (also refer to Fig.~\ref{fig:lip}).  These results demonstrate within the limit of numerical precision that {\em the ground-state 1-RDM spectrum for the lithium atom is pinned to the boundary of the pure set}.

\begin{table}[t!]

  \caption{For the ground state of lithium the pure and Slater distances of the 1-RDM spectrum of natural occupation numbers are shown as functions of the floating-point precision.  While the Slater distance remains constant at $8.53 \times 10^{-5}$ as the floating-point precision is increased, the logarithm of the pure distance decreases linearly with the precision.  These results demonstrate that the ground-state 1-RDM spectrum of lithium is pinned to the boundary of the pure set.}

  \label{t:lip}

  \begin{ruledtabular}
  \begin{tabular}{ccc}
  Precision &  Pure & Slater  \\
  \hline
  5  & $10^{-5}$  & $8.00{\times}10^{-5}$ \\
  10 & $10^{-10}$ & $8.53{\times}10^{-5}$ \\
  15 & $10^{-15}$ & $8.53{\times}10^{-5}$ \\
  20 & $10^{-20}$ & $8.53{\times}10^{-5}$ \\
  25 & $10^{-25}$ & $8.53{\times}10^{-5}$ \\
  30 & $10^{-30}$ & $8.53{\times}10^{-5}$ \\
  35 & $10^{-35}$ & $8.53{\times}10^{-5}$ \\
  \end{tabular}
  \end{ruledtabular}

\end{table}

Table~\ref{t:lipex} shows the Euclidean distances of ground- and excited-state 1-RDM spectra of the lithium atom from the pure and ensemble boundaries and from the Slater point.  Calculations of the 1-RDMs and the Euclidean distances were performed with a numerical precision of thirty decimals. While the spectra of the ground states of a given spin symmetry were always found to be pinned to the boundary of the pure set, {\em the spectra of the excited states were not necessarily pinned}.  For example, the spectrum of excited state~3 lies well within the boundary of the pure set.  The difference in pinning between the ground and excited states was foreshadowed by the discussion in section~\ref{sec:D2} of the necessary 2-RDM condition for pinning.  A 1-RDM spectrum can be pinned to a facet of the generalized Pauli condition only if the 2-RDM is pinned to the boundary of the pure $N$-representable 2-RDM set.  While a ground-state 2-RDM is always on the boundaries of the ensemble and the pure $N$-representable sets of 2-RDMs, an excited-state 2-RDM is not necessarily on the boundary of either the ensemble or pure sets.  Hence, the spectrum of an excited-state 1-RDM is not necessarily on the boundary of the pure set of $N$-representable 1-RDMs.

\begin{table}[t!]

  \caption{For the ground and excited states of the lithium atom, Euclidean distances of the 1-RDM spectra from the pure and ensemble boundaries and from the Slater point are shown.  While the spectra of the ground states of a given spin symmetry were always found to be pinned to the boundary of the pure set, the spectra of the excited states were not necessarily pinned.  For example, the spectrum of excited state 3 lies well within the boundary of the pure set. Calculations of the 1-RDMs and the Euclidean distances were performed with a numerical precision of thirty decimals.}

  \label{t:lipex}

  \begin{ruledtabular}
  \begin{tabular}{ccccccc}
     & \multicolumn{2}{c}{} &  & \multicolumn{3}{c}{Euclidean distance}  \\
    \hline
    State & $S_z$    & Energy (a.u) &       & \multicolumn{1}{c}{Pure}  & \multicolumn{1}{c}{Ensemble}  & \multicolumn{1}{c}{Slater} \\
    \hline
    0     & 0.5   & -7.316 &       & $1.00{\times}10^{-30}$ & $1.00{\times}10^{-30}$ & $8.53{\times}10^{-5}$ \\
    1     &       & -7.230 &       & $1.00{\times}10^{-30}$ & $4.10{\times}10^{-5}$ & $1.41{\times}10^{-4}$ \\
    2     &       & -5.264 &       & $1.00{\times}10^{-30}$ & $1.00{\times}10^{-30}$ & $1.75{\times}10^{-1}$ \\
    3     &       & -5.244 &       & $2.72{\times}10^{-1}$ & $3.65{\times}10^{-1}$ & $8.16{\times}10^{-1}$ \\
    4     & 1.5   & -5.244 &       & $1.00{\times}10^{-30}$ & $1.00{\times}10^{-30}$ & $1.00{\times}10^{-30}$ \\
  \end{tabular}
  \end{ruledtabular}

\end{table}

\subsection{Strongly correlated systems}

\begin{figure}[t!]
  \subfloat[Potential energy surface]{%
  \includegraphics[scale=0.6]{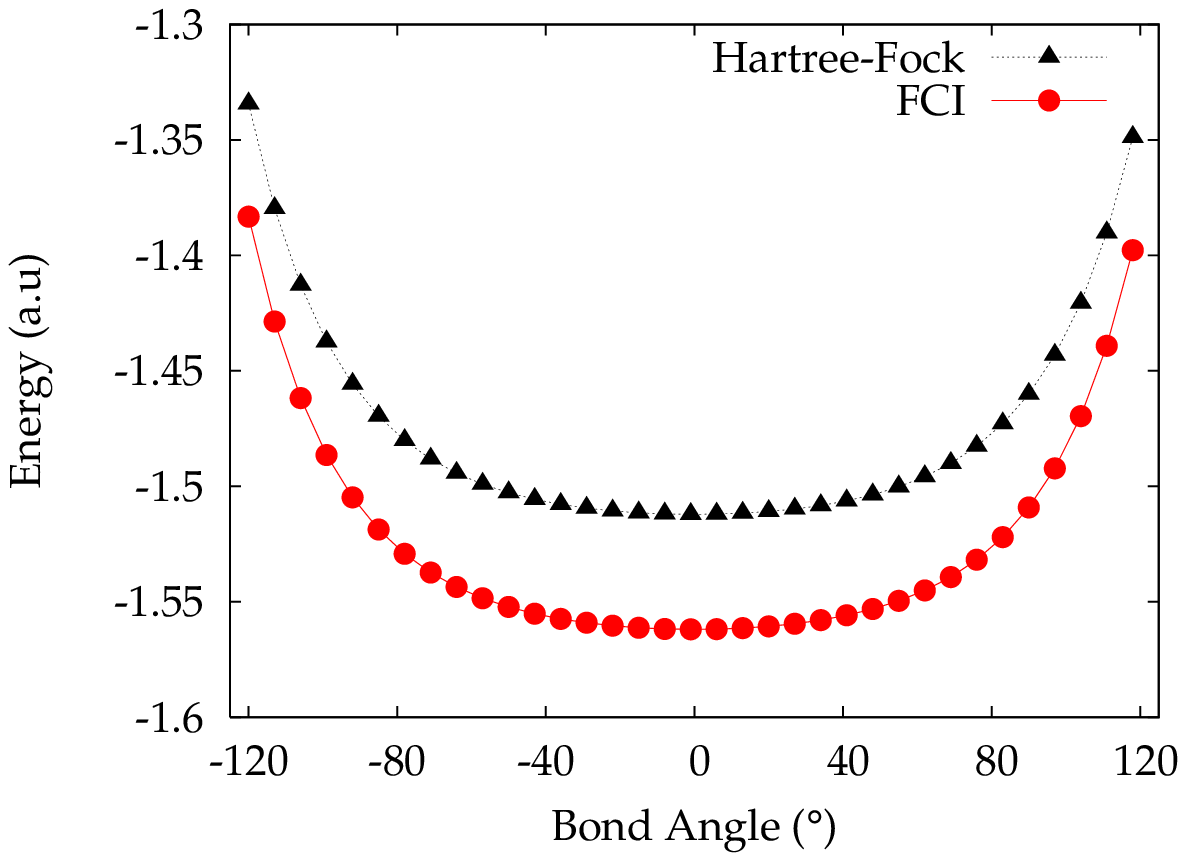}}
  \quad
  \subfloat[Euclidean distance]{%
  \includegraphics[scale=0.6]{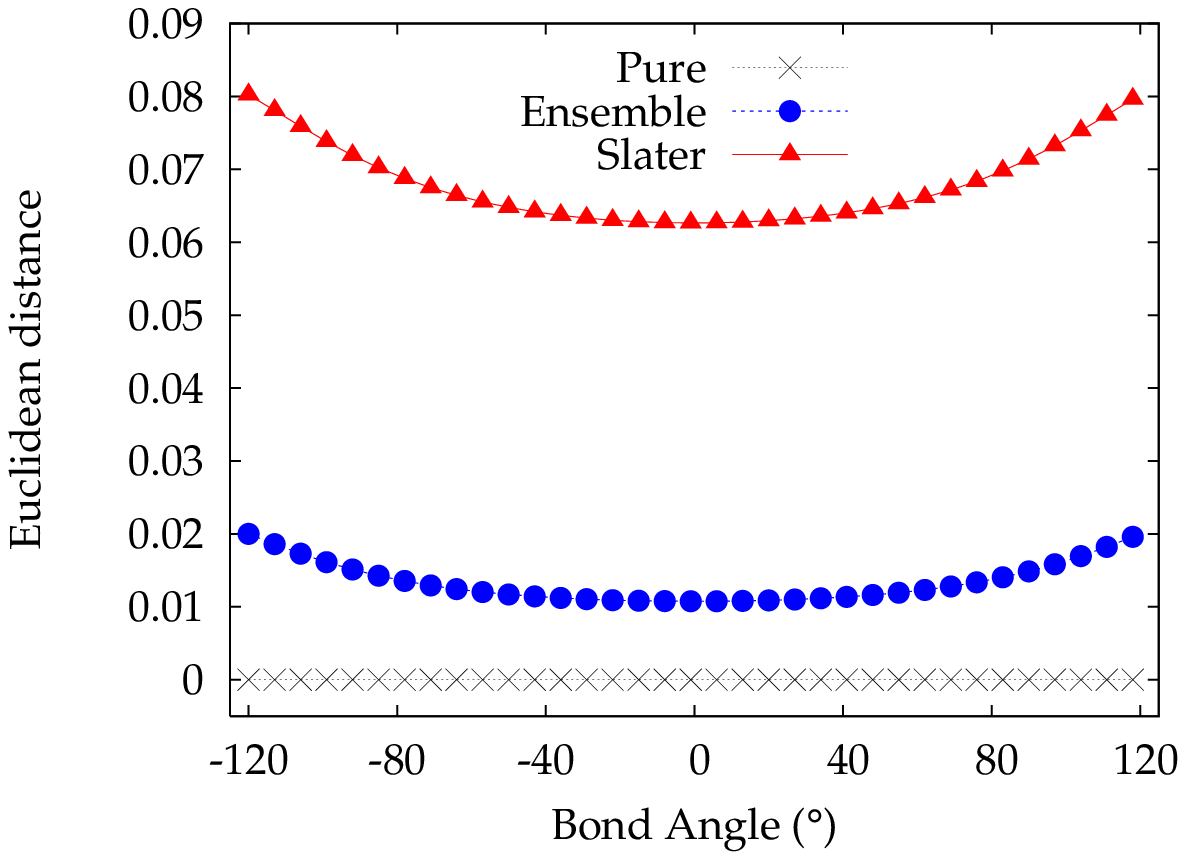}}

  \caption{(Color online) As functions of the bond angle in H\sub3, (a) the Hartree-Fock and correlation energies as well as (b) the minimum Euclidean distances to the boundaries of the pure and ensemble sets and the nearest Slater point are shown.  Both the ensemble and Slater distances in (b) are greater than 0.01 for all bond angles while the pure distance is zero for all angles.  The large distance to the nearest Slater point shows that H\sub3 is significantly correlated.  The vanishing pure distance shows that the generalized Pauli conditions can be saturated even when the traditional Pauli conditions are far from being saturated.} \label{fig:H3}

\end{figure}

The neutral triatomic hydrogen molecule is strongly correlated due to multi-reference effects that arise due to degeneracy in electronic configurations~\cite{K11,JJP12}.  The spectrum of the ground-state 1-RDM, we find, is on the boundary of the pure set for all molecular geometries.  This constitutes the first numerical evidence of pinning of the 1-RDM spectra in a strongly correlated molecule. Table~\ref{tab:3} gives a summary of the results for linear and triangular geometries of H\sub3.  The spectra of the excited-state 1-RDMs are not necessarily on the boundary of the pure set, as demonstrated for both molecular geometries by state~2.  As functions of the bond angle in ground-state H\sub3, (a) the Hartree-Fock and correlation energies as well as (b) the minimum Euclidean distances to the boundaries of the pure and ensemble sets and the nearest Slater point are shown in Fig.~\ref{fig:H3}.  Both the ensemble and Slater distances are greater than 0.01 for all bond angles while the pure distance is zero for all angles.  The large distance to the nearest Slater point shows that H\sub3 is significantly correlated.  The vanishing pure distance shows that the generalized Pauli conditions can be saturated even when the traditional Pauli conditions are far from being saturated.

\begin{table*}[t!]

  \caption{Euclidean distances to the pure and ensemble boundaries and the Slater point for linear and equilateral configurations of neutral triatomic hydrogen H\sub3. The spectrum of the ground-state 1-RDM, we find, is on the boundary of the pure set for both molecular geometries.  The spectra of the excited-state 1-RDMs are not necessarily on the boundary of the pure set, as demonstrated for both molecular geometries by state~2.  The distance to the nearest Slater point, which represents a completely uncorrelated system, shows that the H\sub3 is significantly correlated, especially in the vicinity of the equilateral geometry.}

  \label{tab:3}

  \begin{ruledtabular}
  \begin{tabular}{cccccccccccc}
          &       &       &       &       & \multicolumn{3}{c}{Occupation numbers} &       & \multicolumn{3}{c}{Euclidean distance} \\
    \hline
    Configuration & State & $S_z$    & Energy(a.u) &       & $\lambda_1$    & $\lambda_2$    & $\lambda_3$    &       & \multicolumn{1}{c}{Pure}  & \multicolumn{1}{c}{Ensemble}  & \multicolumn{1}{c}{Slater} \\
    \hline
    Linear & 0     & 0.5   & -2.958 &       & 0.9902 & 0.9789 & 0.9691 &       & $1.0{\times}10^{-30}$ & $1.1{\times}10^{-2}$ & $5.5{\times}10^{-2}$ \\
          & 1     &       & -2.666 &       & 1.0000 & 0.6479 & 0.6479 &       & $1.0{\times}10^{-30}$ & $1.0{\times}10^{-30}$ & $7.0{\times}10^{-1}$ \\
          & 2     &       & -2.448 &       & 0.6667 & 0.6667 & 0.6667 &       & $2.7{\times}10^{-1}$ & $3.7{\times}10^{-1}$ & $8.2{\times}10^{-1}$ \\
          & 3     &       & -2.358 &       & 1.0000 & 0.6530 & 0.6530 &       & $1.0{\times}10^{-30}$ & $1.0{\times}10^{-30}$ & $6.9{\times}10^{-1}$ \\
          & 4     & 1.5   & -2.448 &       & 1.0000 & 1.0000 & 1.0000 &       & $1.0{\times}10^{-30}$ & $1.0{\times}10^{-30}$ & $1.0{\times}10^{-30}$ \\
    Equilateral  & 0     & 0.5   & -3.308 &       & 0.9929 & 0.9909 & 0.9838 &       & $1.0{\times}10^{-30}$ & $7.7{\times}10^{-3}$ & $2.8{\times}10^{-2}$ \\
          & 1     &       & -3.304 &       & 1.0000 & 0.9835 & 0.9835 &       & $1.0{\times}10^{-30}$ & $1.0{\times}10^{-30}$ & $3.3{\times}10^{-2}$ \\
          & 2     &       & -2.652 &       & 0.6667 & 0.6667 & 0.6667 &       & $2.7{\times}10^{-1}$ & $3.7{\times}10^{-1}$ & $8.2{\times}10^{-1}$ \\
          & 3     &       & -2.368 &       & 1.0000 & 0.5182 & 0.5182 &       & $1.0{\times}10^{-30}$ & $1.0{\times}10^{-30}$ & $9.6{\times}10^{-1}$ \\
          & 4     & 1.5   & -2.652 &       & 1.0000 & 1.0000 & 1.0000 &       & $1.0{\times}10^{-30}$ & $1.0{\times}10^{-30}$ & $1.0{\times}10^{-30}$ \\
  \end{tabular}
  \end{ruledtabular}

\end{table*}

\begin{table*}[t!]

  \caption{Euclidean distances of the 1-RDM spectra to pure (Smith), pure (Klyachko), and ensemble (Pauli) sets and the Slater point are shown for the ground states of several four-electron molecules.  Because the ground-state 1-RDM spectra in all cases are pairwise degenerate, they are pinned to the boundary of the Smith set.  The Euclidean distances of the 1-RDM spectra to the pure (Klyachko) and ensemble (Pauli) boundaries are the same in all cases.  The pure (Klyachko) distances are nonzero because the generalized Pauli conditions describing the boundary of the Klyachko polytope break time-reversal symmetry.  All molecules except H$_{4}$ are treated in their equilibrium geometries from the Computational Chemistry Comparison and Benchmark Database~\cite{CCCBD13}.  The H$_{4}$ molecule is treated with the four hydrogen atoms in a square with sides of 1.058~\AA.}

  \label{t:mol}

  \begin{ruledtabular}
  \begin{tabular}{ccccccccc}
  Molecule & \multicolumn{4}{c}{Occupation numbers} & \multicolumn{2}{c}{Pure} & Ensemble & Slater \\
  \hline
          & $\lambda_{1}=\lambda_{2} $    & $\lambda_{3}=\lambda_{4}$    & $\lambda_{5}=\lambda_{6}$    & $\lambda_{7}=\lambda_{8}$           & Smith & Klyachko &       &  \\
  \hline
  LiH   & 1.0000 & 0.9985 & 0.0013 & 0.0002 & 0.00000 & 0.00002 & 0.00002 & 0.00281 \\
  BH    & 0.9991 & 0.9260 & 0.0375 & 0.0375 & 0.00000 & 0.00098 & 0.00098 & 0.12876 \\
  BeH$_2$  & 0.9957 & 0.9938 & 0.0068 & 0.0037 & 0.00000 & 0.00398 & 0.00398 & 0.01527 \\
  H$_4$    & 0.9694 & 0.5000 & 0.5000 & 0.0306 & 0.00000 & 0.03270 & 0.03270 & 1.00187 \\
  \end{tabular}
  \end{ruledtabular}

\end{table*}

Euclidean distances of the 1-RDM spectra to the boundaries of pure (Smith), pure (Klyachko), and ensemble (Pauli) sets and the Slater point are shown in Table~\ref{t:mol} for the ground states of several four-electron molecules in eight spin orbitals ($\wedge ^4[ \mathcal{H}_8 ]$).  Because the ground-state 1-RDM spectra in all cases are pairwise degenerate, they are pinned to the boundary of the Smith set.  The Euclidean distances of the 1-RDM spectra to the pure (Klyachko) and ensemble (Pauli) boundaries are found to be the same in all cases.  The pure (Klyachko) distances are nonzero because the generalized Pauli conditions describing the boundary of the Klyachko polytope break time-reversal symmetry.

\begin{figure}[t!]
  \subfloat[Potential energy surface]{%
  \includegraphics[scale=0.6]{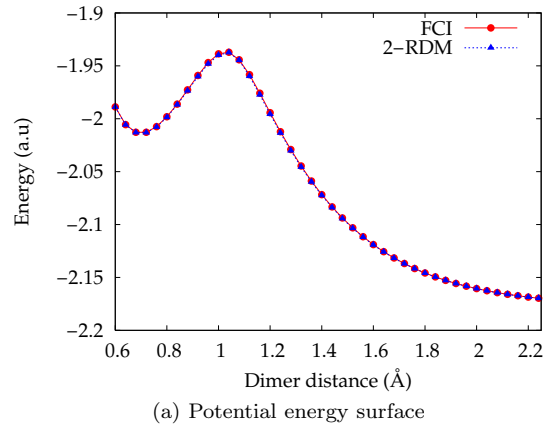}}
  \quad
  \subfloat[Euclidean distance]{%
  \includegraphics[scale=0.6]{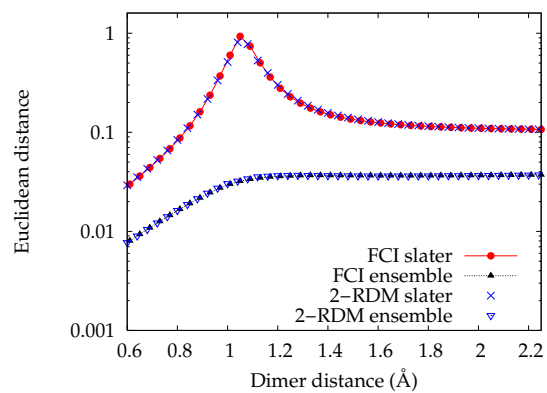}}

  \caption{(Color online) As functions of the distance between H$_{2}$ dimers, the figure shows (a) the potential energy surfaces from FCI and the variational 2-RDM method as well as (b) the minimum Euclidean distances from the 1-RDM spectra to the ensemble set and a Slater point from FCI and the variational 2-RDM method.  The peak in the Slater distance at a dimer distance of 1~\AA\ shows that the maximum electron correlation occurs when the two H$_{2}$ dimers form a square H$_{4}$ molecule.  While the ensemble distance is about 0.01 or larger for all dimer distances,  the 1-RDM spectra are pinned to Smith's set of pure $N$-representable 1-RDMs with time-reversal symmetry.  The figures also show that the FCI and variational 2-RDM methods give similar results for both energies and Euclidean distances.}

  \label{fig:H4}

\end{figure}

The rectangular H\sub4 molecule, comprised of two H\sub2 monomers, has well-documented multi-reference correlations effects in the form of pronounced diradical character~\cite{SM13}.  As functions of distance between H$_{2}$ dimers, Fig.~\ref{fig:H4} shows (a) the potential energy surfaces from FCI and the variational 2-RDM method as well as (b) the minimum Euclidean distances from the 1-RDM spectra to the ensemble set and a Slater point from FCI and the variational 2-RDM method.  The peak in the Slater distance at a dimer distance of 1~\AA\ shows that the maximum electron correlation occurs when the two H$_{2}$ dimers form a square diradical H$_{4}$ molecule.  While the ensemble distance is about 0.01 or larger for all dimer distances,  the 1-RDM spectra are pinned to Smith's set of pure $N$-representable 1-RDMs with time-reversal symmetry.  The figures also show that the FCI and variational 2-RDM methods give similar results for both energies and Euclidean distances.

\section{Discussion and Conclusions}

Generalized Pauli conditions on the 1-RDM spectrum are explored for the ground and excited states of atoms and molecules. We employ Euclidean distance as a systematic means for measuring the distance between 1-RDMs represented as vectors in a Euclidean space~\cite{H78}. While previous work examined the residual of the generalized Pauli conditions to assess their saturation~\cite{PRL13,PRA13}, we compute the minimum Euclidean distances between 1-RDM spectra and facets of the polytope described by the generalized Pauli conditions, also known as pure $N$-representability conditions~\cite{C63, S66, BD72, K06, PRL13, PRA13}.  The Euclidean metric allows us to compare the pure distance with other distances including the minimum distance of the 1-RDM to the boundary of the ensemble $N$-representable set, which we called the ensemble distance, set as well as the minimum distance of the 1-RDM to a Slater 1-RDM, which we called the Slater distance.  Euclidean distances have the following order:
\begin{equation}
{\rm pure} \le {\rm ensemble} \le {\rm Slater}.
\end{equation}
The Slater distance vanishes if and only if the 1-RDM corresponds to a non-interacting Slater-determinant wave function.  Both the ensemble and pure distances can vanish for correlated quantum systems.   The ensemble distance vanishes if and only if the 1-RDM spectrum is pinned to one of the traditional Pauli conditions, and the pure distance vanishes if and only if the 1-RDM spectrum is pinned to one of the generalized Pauli conditions. Importantly, as demonstrated in the examples, the 1-RDM spectra can be on the boundary of the set of pure $N$-representable 1-RDMs even if it is not on the boundary of the set of ensemble $N$-representable 1-RDMs.

The Slater distance provides a measure of electron correlation and entanglement because it vanishes in the absence of electron correlation~\cite{HHH09,HK05,SM13b}.  The pure and ensemble distances, which can vanish in the presence of electron correlation, provide an additional classification of electron correlation and entanglement. In general, the quantum system becomes more correlated the further the 1-RDM is located from nearest Slater point and the boundaries of its pure and ensemble sets.  Proximity of the 1-RDM spectrum to a facet of the pure or ensemble $N$-representable 1-RDM set provides important specific information about the nature and character of the electron correlation and entanglement present in the quantum system.

The ensemble and pure $N$-representable sets of 2-RDMs, we show, provide a necessary condition for the pinning of the 1-RDM to the boundary of the pure $N$-representable 1-RDM set.   Because the 1-RDM is derivable from the 2-RDM through integration over one of the electrons, the 1-RDM can be on the boundary of its pure $N$-representable set only if the 2-RDM lies of the boundary of its pure set.  This necessary 2-RDM condition provides insight into the difference in the 1-RDM between ground and excited states.	Computationally, we find that the 1-RDM spectra from ground states are consistently pinned to the boundary of the pure $N$-representable set while the 1-RDM spectra from excited states are not necessarily pinned to the boundary.  The ground-state 2-RDMs satisfy the necessary condition for the pinning of the 1-RDM to its pure boundary.  In contrast,   the excited-state 2-RDMs do not necessarily satisfy this condition, and in fact, they often lie significantly within the set of ensemble $N$-representable 2-RDMs.

The difference in the 1-RDM spectrum between ground and excited states is significant because it reflects a fundamental difference in the nature of electron correlation between ground and excited states.  Unlike the ground state 1-RDM spectrum, which we find pinned to particular facet of the generalized Pauli conditions, the excited state 1-RDM spectrum can be buried deeper in the polytope, which is associated with greater electron correlation.   Consequently, the geometric difference between ground and excited state 1-RDM spectra may provide a novel explanation, complementary to wave function arguments, for the propensity of excited states to be more correlated than the ground state.

In the 1960s Smith demonstrated that any 1-RDM with time-reversal symmetry and an even number $N$ of electrons is pure $N$-representable~\cite{S66}.  He showed that, if $N$ is even, time-reversal symmetry causes all of the eigenvalues of a 1-RDM to be evenly degenerate.   This degeneracy of the 1-RDM spectra can occur if and only if the 1-RDM is pure $N$-representable.  Here we show that for $N=4$ time-reversal symmetry must be considered to observe pinning of the 1-RDM spectrum from a ground state of an atom or molecule to the boundary of the pure $N$-representable 1-RDM set. For $N = 4$ the 1-RDM spectrum is not pinned to a facet of the polytope defined by the conditions of Klyachko, which do not consider time-reversal symmetry; in fact, for a 1-RDM with even $N$ and time-reversal symmetry, the conditions of Klyachko reduce to the traditional Pauli conditions.  In contrast, all of the 1-RDMs in the Smith set, which obey time-reversal symmetry, are pinned to its boundary.   The presence of symmetry in the quantum state is important in the definition of a system-appropriate set of pure $N$-representable 1-RDMs. Symmetry restricts the set of 1-RDMs to be physically realistic for the set of quantum systems under consideration, and in many cases it is an active constraint in that a state's 1-RDM saturates the symmetry constraint.  In the case of time-reversal symmetry, the symmetry provides sufficiently tight restrictions on the 1-RDM spectrum to guarantee its pure $N$-representability, and thereby highlights an interesting interplay between symmetry and $N$-representability.

In addition to providing a useful measure and classification of electron correlation and entanglement, the generalized Pauli conditions also offer insights into the improvement of electronic structure methods based on wave functions as well as 1- and 2-RDMs.   The theorem of Hohenberg-Kohn~\cite{HK} at the heart of density functional theory~\cite{DFT} intimates that the 1-RDM contains key elements of an atom or molecule's electronic structure including possible signatures for strong electron correlation.   The generalized Pauli conditions further demonstrate that key features of electron correlation and entanglement are encoded within the 1-RDM.  The definition of the set of pure $N$-representable 1-RDMs by the generalized Pauli conditions may provide new insights into the development of practical 1-RDM-based electronic structure methods~\cite{PML10, SDS13}.  The generalized Pauli conditions may also be useful in some cases as further restrictions on the $N$-representability of the 2-RDM~\cite{GP64, E78, CY00, RDM, M12} in variational calculations based on the 2-RDM~\cite{EJ00, NNE01, M02, ZBF04, M04, *M06, CSL06, GM06, HM06, RM08, NBF08, VVV09, S10, GM10, PGG11, M11, B12} rather than the wave function.  Although the generalized Pauli conditions are already implied for even $N$ by the time-reversal symmetry of the 2-RDM, these conditions may be useful for odd $N$ or states without time-reversal symmetry.   As suggested by other authors~\cite{PRL13,PRA13}, the generalized Pauli conditions may also provide insight into the structure of the wave function. Pinning of the 1-RDM to the generalized Pauli conditions provides potentially useful information about the Slater determinants that contribute most significantly to the wave function.

In summary, the generalized Pauli conditions, also known as pure $N$-representability conditions, provide additional constraints beyond the Pauli exclusion principle to ensure that a 1-RDM is derivable from a pure density matrix with a single wave function spectrum.  Using the Euclidean-distance metric, we have explored these conditions for the lithium atom and a variety of molecules at equilibrium and nonequilibrium geometries.  Even with the presence of strong correlation, we find the ground state 1-RDM spectra remain pinned to the boundary of the pure $N$-representable set of 1-RDMs.  For excited states the 1-RDM spectra are not necessarily pinned, and we explain the important difference between ground and excited states through a necessary condition on the 2-RDM for pinning.   The generalized Pauli conditions are useful in the measurement and classification of electron correlation and entanglement.  More generally, these conditions provide fundamental insight into the structure of many-electron quantum systems, which may be useful for both the classification and the computation of strong electron correlation.

\begin{acknowledgments}

DAM gratefully acknowledges the NSF, ARO, and Microsoft Corporation
for their generous support.

\end{acknowledgments}

\bibliography{PinningR1}{}

\end{document}